\documentclass[fleqn,usenatbib]{mnras}

\usepackage{newtxtext,newtxmath}
\usepackage[T1]{fontenc}

\DeclareRobustCommand{\VAN}[3]{#2}
\let\VANthebibliography\thebibliography
\def\thebibliography{\DeclareRobustCommand{\VAN}[3]{##3}\VANthebibliography}

\usepackage{graphicx}
\newcommand{\putpic}[2][1]{\includegraphics[width=#1\linewidth]{plots/#2}}
\newcommand{\nopic}[1]{{\color{red} Image: \detokenize{plots/#1} not found!}}
\newcommand{\pic}[2][1]{ \IfFileExists{plots/#2.pdf}{\putpic[#1]{#2}}{\IfFileExists{plots/#2.png}{ \putpic[#1]{#2} } { \IfFileExists{plots/#2.jpg}{\putpic[#1]{#2}}{\nopic{#2}} } } }

\usepackage{url}

\makeatletter
\def\@to{to}
\makeatother

\usepackage{deluxetable}

\providecommand{\nolinkurl}[1]{\url{#1}}

\usepackage{lipsum}

\title{A high-rate foreground of sub-second flares from geosynchronous satellites}

\author[G. Nir et al.]{
	Guy Nir,\thanks{E-mail: guy.nir@weizmann.ac.il}
	Eran O.~Ofek,
	Sagi Ben-Ami,
	Noam Segev,
	David Polishook,
	Ilan Manulis
\\
Department of Particle Physics and Astrophysics, Weizmann Institute of Science, 76100 Rehovot, Israel
}

\date{Accepted XXX. Received YYY; in original form ZZZ}
\pubyear{2020}

\begin{document}

	\label{firstpage}
	\pagerange{\pageref{firstpage}--\pageref{lastpage}}
	\maketitle
	
	\begin{abstract}
		The Weizmann Fast Astronomical Survey Telescope (W-FAST) is a 55\,cm optical survey telescope 
		with a high cadence (25\,Hz) monitoring of the sky over a wide field of view ($\approx 7$\,deg$^2$). 
		The high frame rate allows detection of sub-second transients over multiple images.
		We present a sample of $\sim 0.1$--0.3\,s duration flares detected in an un-targeted survey 
		for such transients. 
		We show that most, if not all of them,  
		are glints of sunlight reflected off geosynchronous and graveyard orbit satellites. 
		The flares we detect have a typical magnitude of 9--11, 
		which translates to $\sim 14$--16th magnitude if diluted by a 30\,s exposure time. 
		We estimate the rate of events brighter than $\sim 11$\,mag 
		to be on the order of 30--40 events per day per deg$^2$, 
		for declinations between $-20$ and $+10^\circ$, 
		not including the declination corresponding to 
		the geostationary belt directly above the equator, 
		where the rate can be higher. 
		We show that such glints are common in large area surveys (e.g., ZTF and LSST),  
	    and that some of them have a point-like appearance, 
	    confounding searches for fast transients such as Fast Radio Burst counterparts 
	    and Gamma-ray bursts. 
	    By observing in the direction of the Earth's shadow
	    we are able to put an upper limit on the rate of fast astrophysical transients
	    of 0.052\,deg$^{-2}$\,day$^{-1}$ (95\% confidence limit) 
	    for events brighter than 11\,mag. 
	    We also suggest that the single image, high declination flare observed 
	    in coincidence with the GN-z11 galaxy and assumed to be a Gamma-ray burst, 
	    is also consistent with such a satellite glint.
	\end{abstract}


	%%%%%%%%%%%%%%%%%%%%%%%%%%%% INTRODUCTION %%%%%%%%%%%%%%%%%%%%%%%%%%%%%%%%
	
	\section{Introduction}
	
	There are few known astronomical transients with durations lasting less than a second. 
	Some examples include Gamma Ray Bursts (GRBs; \citealt{GRB_review_Zhang_2018}) 
	and Fast Radio Bursts (FRBs; \citealt{FRB_discovery_Lorimer_2007,FRB_counterpart_Lyutikov_Lorimer_2016,FRB_counterparts_multiwavelength_Nicastro_2021}). 
	Stellar objects typically produce flares on time-scales of 
	seconds to tens of minutes \citep{flare_stars_Balona_2015, flare_stars_West_2015}.
	Optical phenomena on sub-second time-scales may include, 
	e.g., FRB counterparts or collisions or outbursts of small objects in the Solar System.
	
	Since only a few optical telescopes observe a large field of view, 
	and also take data at a cadence higher than once-per-second, 
	the lack of detections of such transients is perhaps not surprising. 
	Some notable examples for such surveys include mostly occultation surveys using small telescopes
	\citep{TAOS_survey_Zhang_2013,TAOS_II_camera_Wang_2016,OASES_survey_Arimatsu_2017,Colibri_survey_Pass_2018}
	but also larger telescopes with deeper limiting magnitude, 
	although at a more modest field of view and cadence \citep{Tomoe_Gozen_camera_Sako_2018}.
	These surveys are mostly focused on dips in the lightcurves of bright stars, 
	and not on blind searches for sub-second transients. 
	Some efforts have been made to explore this specific parameter space, 
	but no outstanding discoveries have been made to date 
	(e.g., \citealt{fast_transients_DECAM_Andreoni_2020,FRB_counterpart_search_Andreoni_2020,FRB_counterpart_search_Chen_2020,SGR1935_x_ray_infrared_flares_De_2020}). 

	Large area surveys that are running blind searches for transients
	on a large field of view can detect such sources, 
	but they would be diluted by the long exposure time of the survey, 
	with typical exposure times of $\sim 15-60$\,s, 
	e.g., the Legacy Survey of Space and Time (LSST; \citealt{large_synoptic_survey_telescope_Ivezic_2007})
	or the Zwicky Transient Facility (ZTF; \citealt{Zwicky_transient_facility_Bellm_Kulkarni_2019}). 
	Even if the flare is bright enough to be seen when diluted by a factor of 10--100, 
	it will still be hard to differentiate the flare from asteroids or sensor artefacts, 
	primarily cosmic rays. 
	For this reason, many surveys ignore transients that appear in single images. 
	
	The Weizmann Fast Astronomical Survey Telescope 
	(W-FAST; \citealt{WFAST_system_overview_Nir_2021}) 
	system was designed specifically to explore 
	the parameter space of large field of view and high cadence. 
	The W-FAST main survey takes images 
	with a field of view of $\approx 7$\,deg$^2$ and a cadence of 25\,Hz. 
	This makes it well suited for finding transients 
	lasting between 0.04 and 1\,s, 
	and brighter than 13\,mag. 
	
	We present detections of flares with typical duration of 0.2s and brightness of 9--11\,mag. 
	While some flares' origins remain uncertain, 
	most flares can be definitively associated with short glints from geosynchronous satellites. 
	Since the duration and brightness of all the events in our datasets are similar, 
	it is possible that all such flares are satellite glints. 

	It has long been known that satellite glints (as well as meteors) 
	can produce flashes in naked eye observations and photometer readings
	\citep{flash_background_rate_Schaefer_1987,perseus_flasher_Schaefer_1987}. 
	In this case, however, the flashes are caused by high-orbit satellites, 
	that can appear to be motionless even in arcsecond-resolution images. 
	Recently, \cite{satellite_glints_EvryScope_Corbett_2020} 
	presented a sample of short time-scale glints. 
	Their sample is larger than that presented here, 
	allowing them to better characterize the magnitude distribution 
	of flares at different regions of the sky. 
	In the equatorial region we find an order of magnitude higher rate, 
	including fainter glints that are most likely not detectable in their survey. 
	Furthermore, their integration time is 120\,s 
	while our short exposures allow us to measure the durations 
	of the events directly. 
	\cite{geosats_DebrisWatch_Blake_2020} have studied 
	satellites, spent rockets and smaller debris at geosynchronous altitude, 
	showing the existence of a population of objects with typical sizes of 
	tens of cm or smaller, 
	that are often tumbling or rotating. 
	These object are good candidates for creating
	the glints we present in this work. 

	We argue that sub-second time-scale glints, 
	particularly those of high orbit satellites, 
	are a common foreground to synoptic surveys. 
	These glints should be considered when discussing
	survey strategies such as splitting a single integration into several exposures. 
	
	In Section~\ref{sec: observations} we present the data acquisition and the observational sample, 
	in Section~\ref{sec: results} we present the results of our analysis of a set of flares and their properties, 
	and we explain the nature of point-source flares in our data. 
	In Section~\ref{sec: other surveys} we discuss the implications of such short flares 
	on large area surveys searching for fast transients, 
	and in Section~\ref{sec: conclusions} we summarize our conclusions. 
	
	\section{Observations}\label{sec: observations}
	
	We used the W-FAST system, 
	which spends most of its time taking high cadence (25\,Hz) 
	images of dense fields.
	The system has a pixel scale of 2.29\,$''$\,pix$^{-1}$, 
	and typical seeing (atmospheric and instrumental) of $\approx 4''$. 
	The system is fully described by Nir et al (in prep). 
	Observations reported in this work were
	obtained using the F505W filter 
	(rectangular band pass between 400 to 610\,nm). 
	During these observations, the full-frame images are not saved, 
	as the data rate of about 6\,Gbits s$^{-1}$ is too large to store continuously\footnote{
	       The data rate is approximately twice that of the planned LSST, 
           assuming it produces 3.2 Gigapixel, 16-bit images every 15 seconds, 
           and twice of that of the the Large Array Survey Telescope \citep{LAST_telescope_Ofek_Ben_Ami_2020}. 
    }
	using our current hardware. 
	While it is possible to save these data continuously, 
	doing so for long periods of time 
	requires a substantial increase in the volume
	of the storage hardware that we have on-site. 
	
	The camera produces batches of one hundred images every four seconds, 
	and we produce cutouts and stack images from these full frame images, 
	which are then discarded. 
	The cutouts include $15\times 15$ pixels around $\approx 2000$ stars, 
	and the stack images include the full sensor data, but on a cadence of 4\,s. 
	
	In addition to cutouts around stars we also run a custom algorithm that,
	in real time, 
	identifies local maxima in the entire image. 	
	We begin by masking the stars and any other constant light sources in the field.
	We input the masked images into the detection algorithm.
	To make the detection process fast enough to run in real time 
	over the full-frame data set, 
	we trigger on a single pixel surpassing 256 ADU. 
	This excludes most noise sources 
	(with typical per-pixel rms of $\sim 3$ ADU)
	and allows us to scan only the most significant byte out of the
	two bytes representing each pixel. 
	The high threshold is not adjusted for observing conditions 
	such as seeing, background or airmass, 
	and the detection pipeline does not account 
	for the light spread outside the brightest pixel. 
	This makes for a very inefficient detection method, 
	which is designed for speed rather than sensitivity. 
	The threshold of 256 ADU corresponds to an average 
	magnitude of $\approx 10$, depending on observing conditions. 

	When peaks are identified, a cutout around each peak is saved, 
	including a $15\times 15$ stamp around the peak location, 
	for each of the one hundred frames in that batch. 
	These cutouts are then checked for flares lasting more than one frame, 
	and those are saved for further processing. 
	The majority of peaks that are detected are cosmic rays, 
	which hit the sensor at a rate of $\approx 0.5$ s$^{-1}$. 
	These appear in a single frame and thus are not saved. 
	Since we began collecting flares in a systematic way 
	we have detected a few each night. 
	
	We present a sample of the observations taken during 2020 August-September. 
	There are a number of fields observed either on or off the ecliptic plane. 
	We disqualify runs where data quality was poor, e.g., if the software crashed
	and did not save the flare data as expected. 
	A summary of the observing runs is shown in Table~\ref{tab: observation log}. 
	
	\section{Results}\label{sec: results}
	
	\subsection{Removing Low Earth Orbit satellites}
	
\begin{figure}
		
		\centering
		
		\pic[0.8]{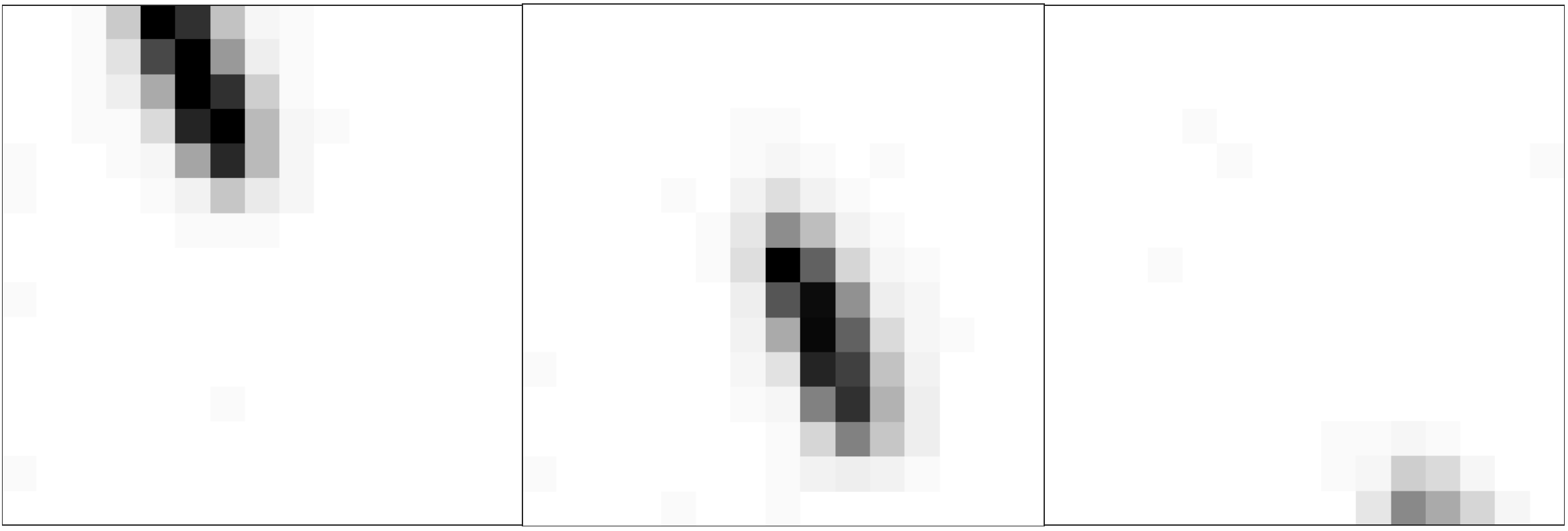}
		
		\caption{An example of a low Earth orbit satellite. 
			The PSF of each individual image is clearly streaked, 
			and the motion of the object between frames is apparent. 
			The (inverted) scale for each image is the same, 
			with the peak pixel value of 763 ADU in the first frame
			and 616 ADU in the central frame. 
		}
		\label{fig: leo sat example}
		
		% this figure was produced by scripts/geosat_summary.m
		
	\end{figure}

	Some of the flares detected by our system were clearly Low Earth Orbit (LEO) satellites. 
	These move at hundreds of arcseconds per second and are visibly streaked 
	even in the high-cadence data. 
	An example for such a streak is shown in Figure~\ref{fig: leo sat example}. 
	
	We disqualify such streaks using hypothesis testing that is done 
	by comparing the Signal to Noise ratio (S/N) 
	calculated in two ways: 
	(a) the S/N assuming the object is a point source, 
	    using a standard matched-filter \citep{matched_filter_Turin_1960} 
	    with a gaussian kernel that has a width of $\sigma_g=1$ pixel;
	(b) the S/N assuming the object is streaked, 		
	    which we find using the the Radon transform map $S_R$
	    of the above-mentioned, filtered image. 
	    The Radon transform sums the pixels along lines 
	    of different angle and starting position  \citep{Radon_transform_1917,Radon_transform_1986}.
	    To normalize the sum of pixels to the noise from those pixels, 
	    we also calculate the Radon transform of a uniform map, $V_R$, representing the background variance, 
	    and divide the two maps to get 
	    
	    \begin{equation}
	    	(\text{S/N})_R = \frac{S_R}{\sqrt{V_R}}. 
	    \end{equation}
	    
	    The maximum of $(\text{S/N})_R$ represents the signal-to-noise ratio of a streaked object. 
		This method has been shown to be optimal for detection 
		of streaks in background dominated images \citep{streak_detection_Nir_2018}. 		
	In both cases we first subtract the median of the cutout to try and reduce the background. 
	To increase the S/N of short streaks, 
	we first crop the image to the central 7 pixels around the brightest point, before using the Radon transform. 
	A simple hypothesis test \citep{optimal_statistic_lemma_Neyman_Pearson_1933}
	checks if the Radon S/N is higher than the point source S/N by a factor\footnote{
		To accommodate the width of the PSF, which is equivalent to a short streak of width 1--2 pixels, 
		we increase the threshold for identifying an object as a point source 
		by a factor of $\sqrt{2}$, which is the expected increase for a two-pixel streak. 
	}
	 larger than $\sqrt{2}$, 
	and if so, we assume it is a streaked source and do not include it in the sample. 
	Furthermore, we disqualify flares occurring less than 6 pixels from the edge of the sensor, 
	since such positions place a large part of the cutout outside the sensor region, 
	and it becomes harder to determine the nature of the flare.

	\subsection{Repeating point-like flares}
	
\begin{figure}
		
		\centering
		
		\pic[0.8]{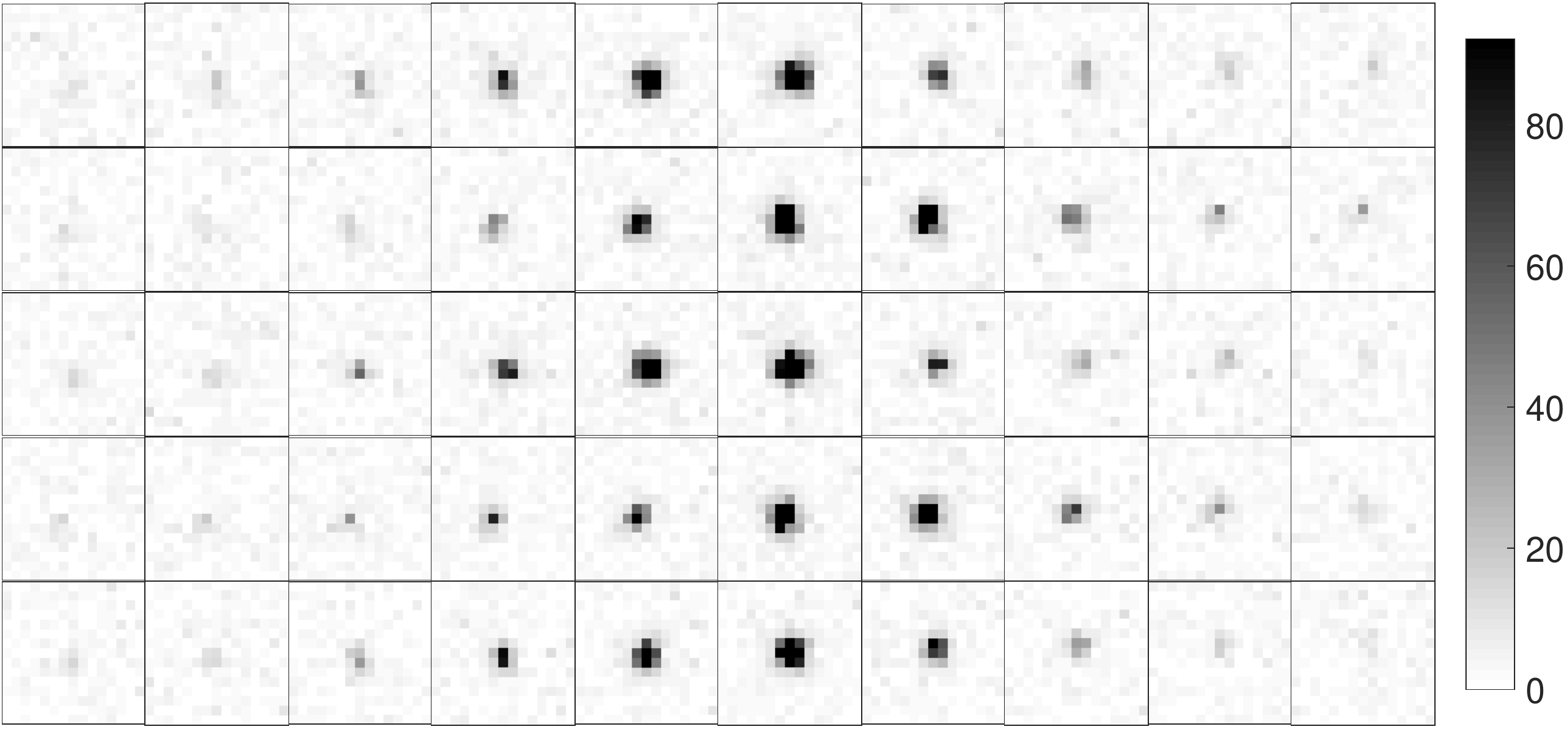}
		\caption{Examples for five glints taken at different locations in the same field of view, 
	 			within the time span of a few minutes. 
 				Each row is for a separate flare, 
 				showing ten consecutive frames around the time of the peak. 
 				The images are all set to the same (inverted) scale, 
 				that shows the range of brightnesses as the flare appears and fades. 
 				The flares have a typical time-scale of 3--5 frames, or 0.12--0.2\,s. 
 			}
		\label{fig: geosats example flares}
		
		% this figure was produced by scripts/geosat_summary.m
		
	\end{figure}

	In Figure~\ref{fig: geosats example flares}
	we show the shape and temporal behavior of 
	a set of point-like flares, detected within a few minutes of each other. 	
	Each row in the Figure is a separate flare, 
	showing that each event brightens and fades in about 0.2 seconds. 
	The shape of each image hotspot is consistent with the system PSF. 
	The images were collected on 2020 August 6th, around 22:38:00 (UTC)
	within a field centered on right ascension of $22^h04^m01.3^s$ 
	and declination of $-10^\circ 29^{'}54.7^{''}$ (J2000).
	Each flare appears to be stationary in the images, 
	ruling out objects in LEO that appear streaked 
	even in single frames. 
	
\begin{figure}
		
		\centering
		
		\pic[0.8]{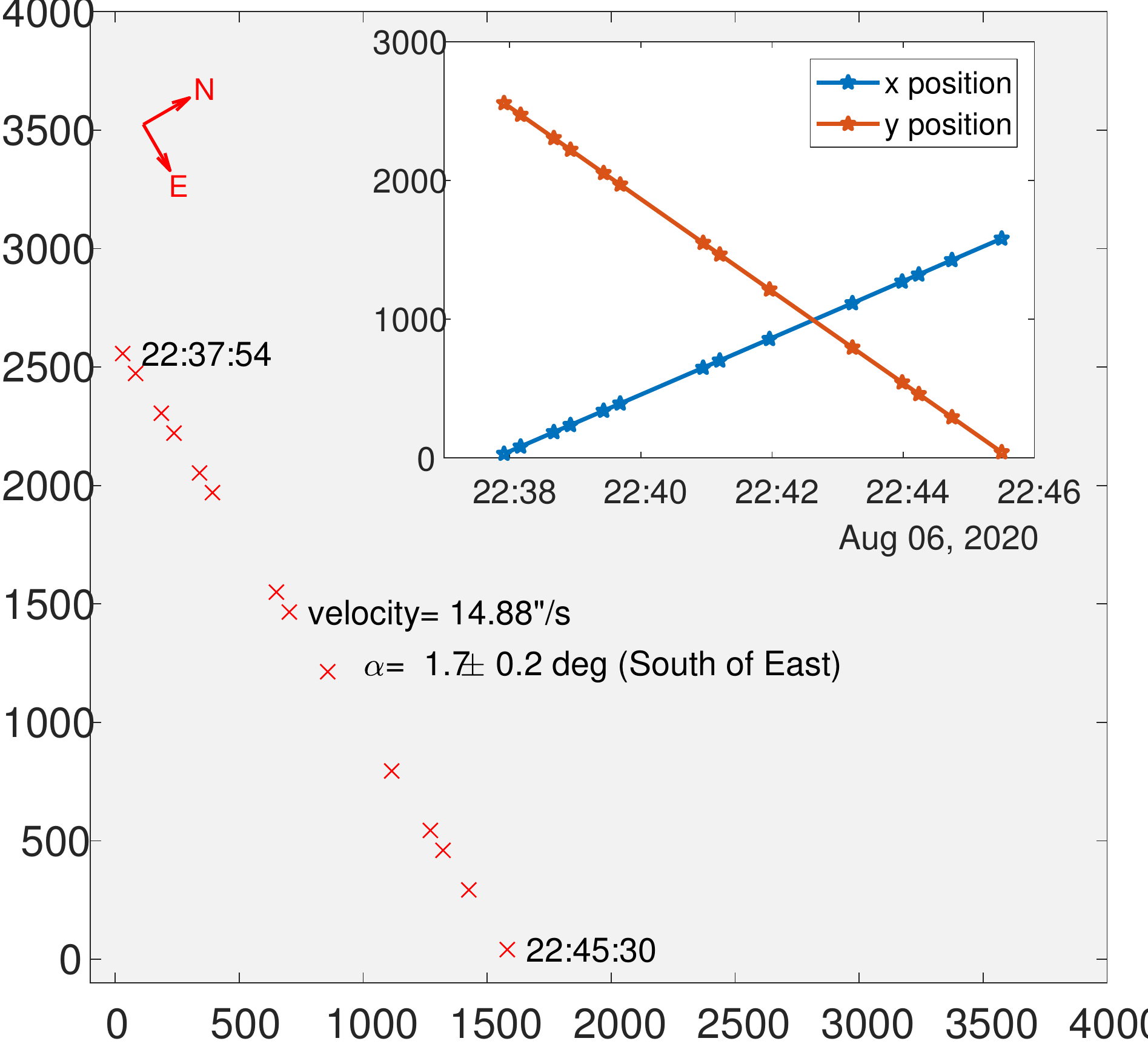}
		\caption{Position of repeating flares from the same object as shown in Figure~\ref{fig: geosats example flares}. 
			The larger plot shows the position of the object at each point on the sensor where a flare occurred. 
			We associated 14 different flares with this object, all aligned in a straight line. 
			The start and end points are marked with the UTC times of the first and last flare in this set. 
			The inset shows the $x$ and $y$ position of the object as a function of time. 			
			The velocity of the object is measured to be close to $14.88''$ per second. 
			The object's velocity's position angle is $91.7\pm 0.2^\circ$ (Measured from North towards East).  
			%			The direction of the object is a few degrees South of East. 
			The time delay between flares is not constant. 
			Clearly the flares are coming from a geosynchronous satellite. 
		}
		\label{fig: geosat trajectory}
		
		% this figure was produced by scripts/geosat_summary.m
		
	\end{figure}

	The example set of flares in Figure~\ref{fig: geosats example flares} 
	shows five flares out of 14, detected in the same field of view, 
	within a time frame of a few minutes. 
	The flares do not repeat at the same location, 
	moving hundreds of arcseconds during tens of seconds between flares,
	spread along a straight line. 
	We thus assume these flares are associated with a single, repeating object. 
	Of course the rate of motion disqualifies the object from being outside the Solar System. 
	Measuring the position of the flares in time across the field of view, 
	as shown in Figure~\ref{fig: geosat trajectory},
	gives the velocity of the source to be 14.88 arcsec s$^{-1}$, 
	at an angle of $\approx 2^\circ$ South of the East direction, 
	consistent with what we expect from geosynchronous satellites. 
	The majority of flares repeating in a single field 
	can be grouped in this way into repeating objects 
	moving with similar rates along a straight line. 
	We therefore conclude that this repeating flare,  
	and most likely all similar sources in our sample, 
	are in fact glints from geosynchronous satellites. 
	
	In this specific example the time delay between flashes are 
	15.2, 30.4, 15.2, 30.3, 15.2, 75.9, 15.2, 45.5, 75.8, 45.5, 15.2, 30.3, 45.5\,s, 
		repeating in multiples of 15.2\,s, where the multiplication is not in any particular pattern. 
	The duration of each flare, taking 3--5 frames to appear and disappear, 
	can be translated to the instantaneous angular velocity, 
	if we model the flashes as coming from a flat, mirror-like surface:
	\begin{equation}
		\omega \simeq \frac{D_\text{ang,sun}}{2 N_\text{frames} \Delta T} = 1.56_{-0.31}^{+0.52} \text{ deg s}^{-1}, 
	\end{equation}
	where $D_\text{ang,sun}$ is the sun angular size, in degrees ($0.5^\circ$), 
	$N_\text{frames}$ is the number of frames where the flash is seen, 
	and $\Delta T$ is the time between frames. 
	The angular frequency and its errors are calculated by setting $N_\text{frames}=4\pm 1$. 
	This translates to a period of $P=230.4\pm 57.6$\,s. 
	The fact that this period is longer than the typical time between 
	flashes suggests there is more than one reflective surface, 
	at different positions on the satellite. 
	It is important to note that our sample includes
	several satellites with short intervals between flares, 
	and others that show a single flare that does not repeat. 
	The duration of flares is also not constant and lasts between 
	one to seven frames.\footnote{
		Most flares that appear in a single frame are excluded by our detection algorithm. 
		However, some single frame flares can be detected, 
		if multiple such flares, separated by a short interval, 
		appear in the same 100 frame (4 second) batch.
		We see several cases where such rapid repeating 
		flares each appear in a single frame. 
		It may be that other single frame flares 
		exist in the data but are not detected 
		by our specific algorithm. 
}
	This suggests there are many rotation periods and 
	many geometries for the reflecting surfaces on the satellites. 
	
	\subsection{Magnitude distribution}
	
\begin{figure}
		
		\centering
		
		\pic[0.8]{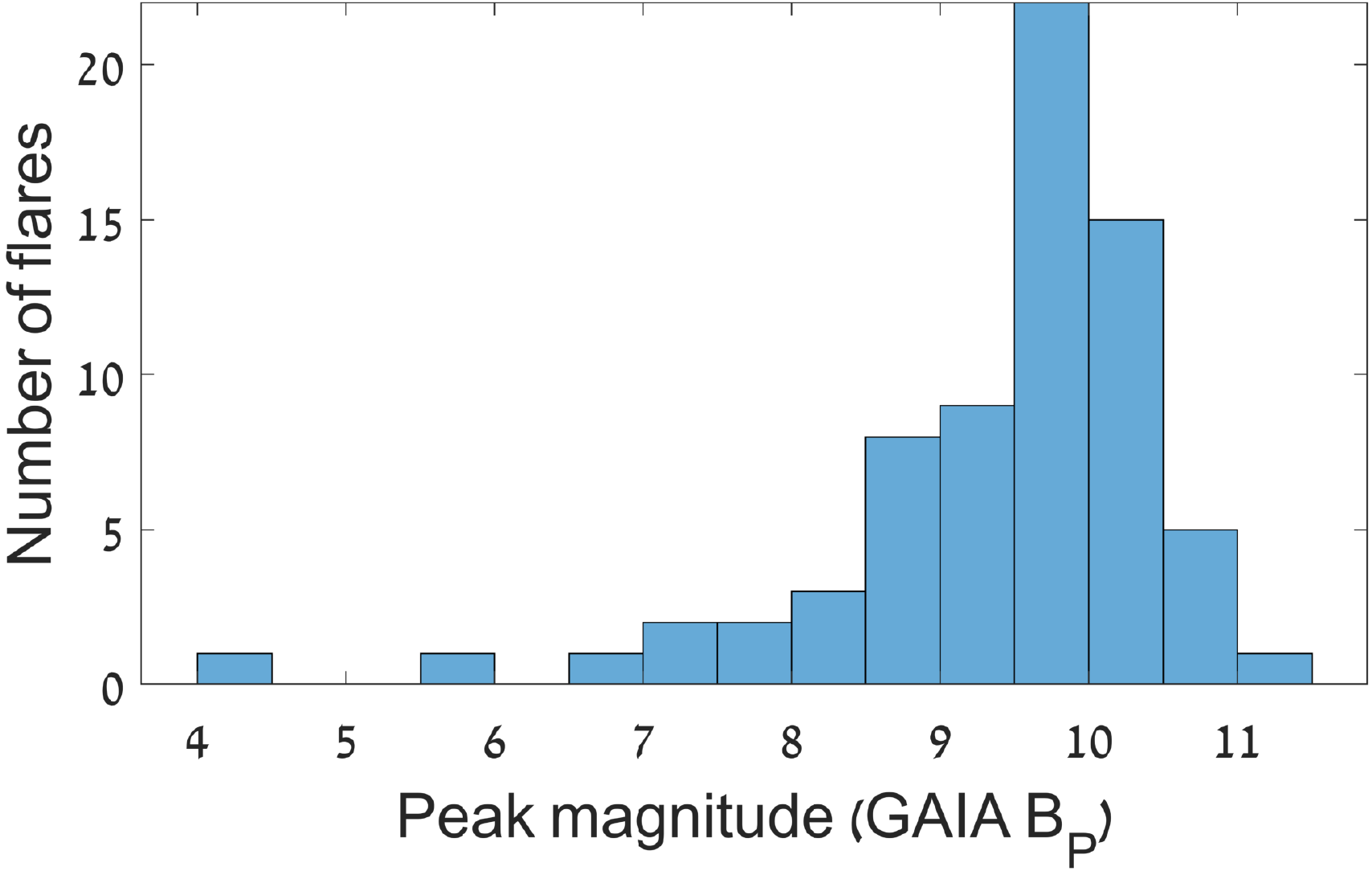}
		
		\caption{Peak magnitude of glints detected in this sample. 
			The flux of each glint is compared, at peak, 
			with the current image zero-point, 
			based on the magnitudes and fluxes of other stars 
			in the image. 
			The stars' magnitude is matched to the GAIA DR2 $B_P$ filter. 
			For objects with multiple glints we plot the brightest magnitude. 
			The scatter for multiple glints is typically less than one magnitude. 
		}
		\label{fig: geosat magnitude histogram}
		
		% this figure was produced by scripts/geosat_summary.m
		
	\end{figure}

	For each flare we calculate the flux using an aperture of 3 pixels radius ($6.87''$)
	that is centred on the position of the flare. 
	We subtract the background using an annulus 
	with inner and outer radii of 7.5 and 10 pixels (17.2 and $23''$). 
	The flux for each frame is compared to the image zero point calculated against other stars 
	using the same photometric pipeline. 
	The stars' magnitudes are compared to the GAIA $B_P$ filter data from DR2 \citep{GAIA_data_release_two_2018}, 
	using tools developed by \cite{matlab_package_Ofek_2014} and \cite{catsHTM_Soumagnac_Ofek_2018}, 
	with astrometric solutions given by \cite{astrometry_code_Ofek_2019}.
	The flares commonly have a magnitude of 9--11 in this band. 
	A histogram of the peak magnitude of all flares is shown in Figure~\ref{fig: geosat magnitude histogram}.
	For repeating flares we include the maximum magnitude 
	of all repeated flares (see discussion below). 
	Repeated flares typically have a magnitude scatter less than 1\,mag. 
	
	The mean magnitude of the peaks of the flares 
	in the example discussed above is around 9.5 in GAIA $B_P$, 
	likely due to the high detection threshold we used to discover them. 
	If we model these flashes as reflections on a smooth, mirror-like, circular surface, 
	we can estimate the size of the reflector based on the measured flux. 
	The magnitude of the flash would be given by
	\begin{equation}
		M_\text{ref} = M_\text{sun} - 5 \log10 \left(\frac{D_\text{ang,ref}}{D_\text{ang,sun}} \right) -2.5 \log10(A), 
	\end{equation}
	where $M_\text{sun}=-26.74$ is the Sun's apparent magnitude,
	$A$ is the mirror albedo (fraction of flux reflected), 
	and $D_\text{ang,ref}$ is the reflector angular diameter. 
	We assume the satellite is on a circular orbit. 
	It was observed at an elevation angle of $47.9^\circ$,
	which gives a slant range of $R=37,200$\,km. 
	Using this distance we can estimate the reflector's physical size:
	\begin{equation}
		D_\text{ref} = R_\text{ref} D_\text{sun,ang}^\text{(radians)} 10^{(M_\text{ref}-M_\text{sun})/5}/\sqrt{A}. 
	\end{equation}
	For $M_\text{ref}=9.7$ we get $D_\text{ref}= 1.8/\sqrt{A}$ in cm. 
	If $A=0.04$, typical of glass windows, 
	we get a diameter of 9cm. 
	
	\subsection{Declination distribution}
	
	We observed for 119.40 hours, 
	found 1341 LEO satellites, 
	and 862 point-source flares, 
	which we associate with 76 
	distinct geosynchronous satellites. 
	
\begin{figure}
		
		\centering
		
		\pic[0.8]{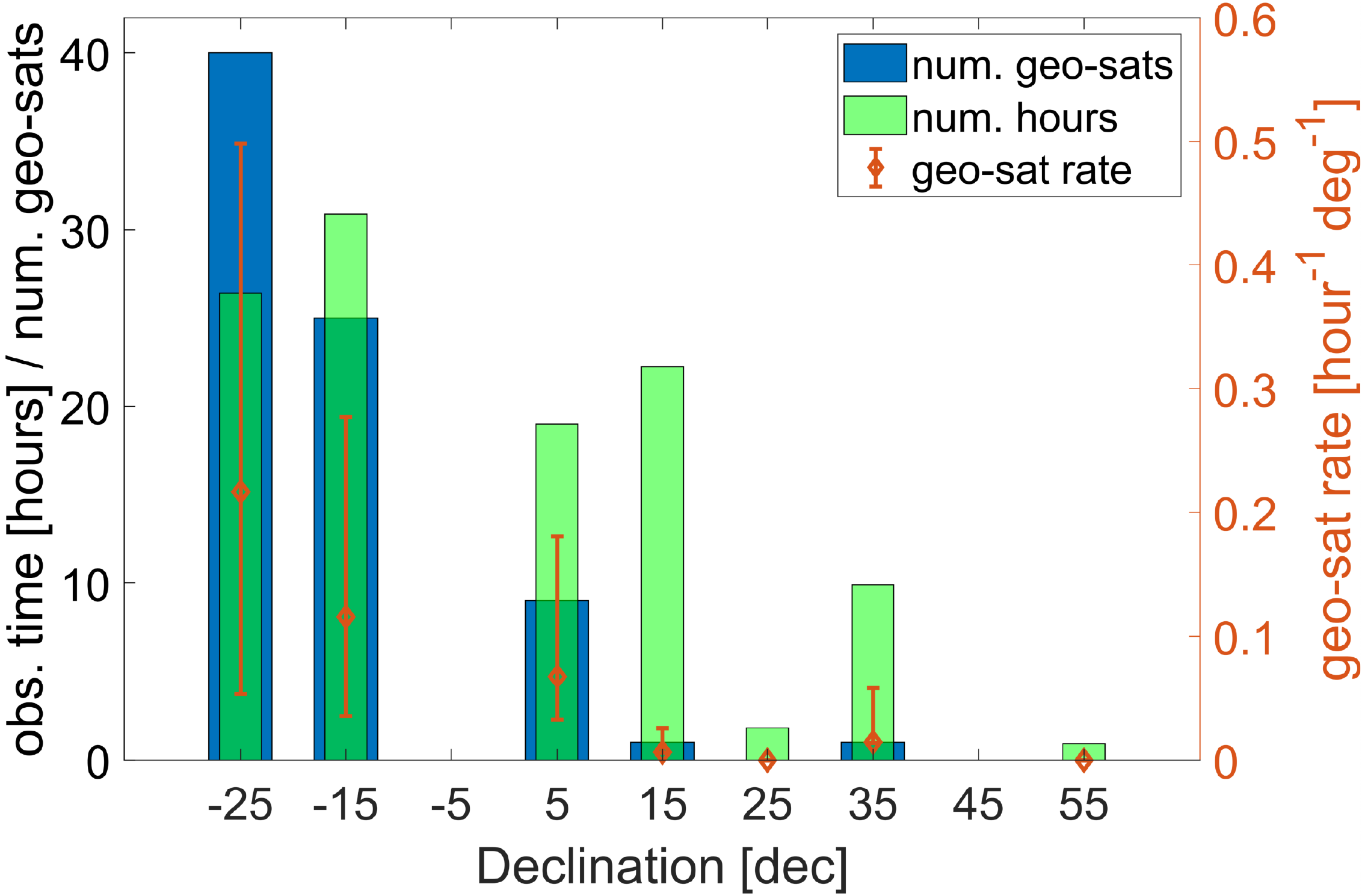}
		
		\caption{The number of hours and the number of geosynchronous satellites 
			     as a function of declination. 
			     Each satellite is comprised of one or multiple flares that were grouped
			     based on velocity that is consistent with geosynchronous orbit. 
%			     Fields with declination $<30^\circ$ are all close to the ecliptic plane ($<5^\circ$).			     
		}
		\label{fig: geosat declinations}
		
		% this figure was produced by scripts/geosat_summary.m
		
	\end{figure}

	In Figure~\ref{fig: geosat declinations}, 
	we plot the number of flaring objects as a function of the field declination. 
	We do not count fields where the Earth's shadow intersects the line of sight
	to the coordinates of the field centre, at the orbital radius of geosynchronous satellites.
	In those fields, we did not detect any flares, 
	but we do not include them in the rate calculations as satellites are not expected to be seen.
		  
	To calculate the number of individual objects that often have multiple flares, 
	we cluster the flares in each run. 
	The criteria for clustering two flares into a single source were that 
	\begin{itemize}
		\item The flares occur less than 600\,s apart in time. 
		\item The flares are close enough on the sky to come from an object moving with a speed of up to\footnote{
			The clustering was based solely on the distance and time between flares.
			We did not require the flares to be on a straight line or to have similar magnitudes. 
			Since glinting objects rarely appear at the same time and nearby on the sky, 
			it is unlikely that flares would be associated with the wrong object during the clustering. 
		} $15''$ s$^{-1}$. 
	\end{itemize}
	 
	In Table~\ref{tab: observation log} we specify for each run the number 
	of individual objects and the average number of flares per object in that run, 
	and whether that run is inside the Earth's shadow at the radius of geosynchronous satellites. 
	The number of satellites and the number of hours per run, 
	as a function of declination, is shown in Figure~\ref{fig: geosat declinations}. 
	The blue bars represent the number of hours observed in each $10^\circ$ declination bin, 
	while the green bars show the number of satellites detected. 
	These include only satellites moving with a velocity 
	consistent with geosynchronous orbit. 
	The rates are given as red diamonds, 
	with the error bars showing the 90\% confidence intervals 
	\citep{poisson_confidence_Gehrels_1986}. 
	
	We see more satellites on low declinations ($-20$ to $+10^\circ$), 
	which is reasonable as many geosynchronous satellites spend 
	much of their time close to the equatorial plane. 
	A very rough estimate of the lower limit on the rate of such glinting objects
	can be calculated from the number of observed objects and the observing time. 
	We find the rate of individual flares, brighter than $\sim 11$\,mag 
	to be $1.73^{+0.17}_{-0.16}$ deg$^{-2}$ hour$^{-1}$ 
	at a declination $-25^\circ$, 
	and a somewhat lower rate of $1.26_{-0.16}^{+0.17}$ deg$^{-2}$ hour$^{-1}$ 
	at a declination of $+5^\circ$. 
	The uncertainties account for 90\% confidence on the Poisson distribution, 
	but do not account for systematics as discussed below. 
	If we count individual objects regardless of the number of repeat flares it displays, 
	the rate is decreased by an order of magnitude, 
	giving a rate of $0.22^{+0.07}_{-0.05}$ deg$^{-2}$ hour$^{-1}$ at declination $-25^\circ$ 
	and $0.07^{+0.05}_{-0.03}$ deg$^{-2}$ hour$^{-1}$ at $+5^\circ$. 
	
	These rates do not include geostationary satellites that are always 
	on the equatorial plane.	 
	From our observatory at latitude 30.59$^\circ$N, the geostationary orbit
	is seen at an apparent declination of $-5^\circ$. 
	Unfortunately, we did not observe any fields in this declination, 
	so we cannot constrain the rate on the geostationary belt directly above the equator. 
	
	This small survey is not complete when it comes to counting satellites:
	the threshold for detection of strong peaks is constant, 
	and set to a pixel value of 256 ADU, 
	without considering changes in seeing, background, or airmass. 
	The detection algorithm requires the signal in at least two frames 
	to be five times brighter than the background,
	which excludes most flares lasting a single frame. 
	Furthermore, some of the flares may have been counted as streaked
	if the seeing was particularly bad. 
	Finally, the survey does not sample a wide range of 
	coordinates on the sky in a systematic way, 
	instead the survey strategy was determined 
	by other considerations for other science goals. 
		
	During these observations, we have taken images of fields inside Earth's shadow 
	(at the radius of geosynchronous satellites)
	for a total of 8.2 hours.
	In this direction no flares were detected,
	which gives an upper limit on the rate of astrophysical
	flares brighter than 11th magnitude of $0.052$ deg$^{-2}$ hour$^{-1}$ (95\% one-sided confidence limit). 
	
	\section{Implications for other surveys}\label{sec: other surveys}
	
	We consider the appearance of these glints in large area surveys, 
	such as ZTF \citep{Zwicky_transient_facility_Bellm_Kulkarni_2019} and LSST \citep{large_synoptic_survey_telescope_Ivezic_2007}.
	While satellites usually appear clearly streaked 
	in long exposures (e.g., more than 1 second), 
	the satellite's diffuse reflection may be too dim to detect even in large telescopes,  
	and only the short-duration flash would be seen.
	If the object rotates quickly, 
	the duration of the flash may become shorter than 0.05s, 
	and the streaked nature of the flash could be smaller than $0.75''$. 
	Under typical seeing conditions, 
	such a glint could have a similar shape as that of point sources in the image. 

	If the exposure is longer than the flash duration, 
	the effective magnitude of the flash would be linearly diluted by the exposure time:
	\begin{equation}
		M' = M + 2.5 \log\left(\frac{\Delta t}{T}\right),
	\end{equation}
	where $M$ is the native magnitude 
	(that would be measured using an exposure time equal to the event duration),  
	$\Delta t$ is the duration of the flash, 
	and $T$ the survey exposure time. 
	For a survey like ZTF, with an exposure time of 30s, 
	the flashes could be diluted by a factor of more than 100, 
	giving an effective magnitude of 15 or dimmer. 
	These flashes would appear in a single image and would 
	be very hard to discern from astrophysical sources. 
	
	Since ZTF does not take consecutive images of the same field back to back \citep{Zwicky_transient_facility_Bellm_Kulkarni_2019}, 
	it is impossible to detect multiple flashes from the same satellite. 
	Even if it were to implement such an observing strategy, 
	as is considered for the upcoming LSST, 
	there is no guarantee that the flash will repeat in both frames. 
	On the other hand, in a fraction of the cases, multiple flashes
	would appear on the same image. 
	This would probably be rare, as the typical repeat time delay
	we have seen is not much shorter than the ZTF exposure time. 
	Therefore it will not be possible, in general,  
	to identify the nature of these glints 
	based on repeating flashes. 
	
	When observing low declinations (around $20^\circ$ from the equatorial plane), 
	the detection rate for ZTF would be $\gtrsim 0.5$ flashes in each 47\,deg$^2$ image, 
	at least hundreds of events per night. 
	This is a lower limit, 
	as fainter flashes likely exist, 
	and their rate could be orders of magnitudes larger. 
	This population of fainter flashes would not be seen by our survey, 
	but could be detectable by ZTF. 
	In order to fully understand the implications, 
	deeper surveys for fast satellite glints are needed. 
	
%	As a comparison to the rate we present here, 
%	each ZTF image contains hundreds or thousands of detections, 
%	astrophysical sources and image artefacts. 
	Most sources in ZTF images are constant over multiple images, 
	but asteroids and image artefacts such as cosmic rays have a considerably 
	higher rate than the glint rate presented here. 
	Main belt asteroids could be as common as $\approx 30$\,deg$^{-2}$ for $R<19$ (1400 in the ZTF field), 
	but can be as high as $\approx 700$\,deg$^{-2}$ for deeper surveys up to $R\sim 25$
	\citep{faint_asteroid_density_DECAM_Heinze_2019}.
	Off the ecliptic the number of main belt asteroids becomes sub-dominant.
	The rate of cosmic ray hits, $\sim 15$\,deg$^{-2}$ (700 in the ZTF field)
	also contributes many single-image detections. 
	However, a significant fraction of cosmic rays 
	can be differentiated from astronomical sources
	using various methods (e.g., \citealt{cosmic_ray_detection_van_Dokkum_2001,image_subtraction_ZOGY_Zackay_2016}).
	Random noise fluctuations, 
	assuming a $5\sigma$ detection threshold, 
	would contribute $\mathcal{O}(10)$ false detections for each square degree 
	(50-200 in the full field of view of ZTF). 
	While the satellite glints at the magnitudes we have probed 
	have a much lower rate than 
	these single-image sources, 
	the satellites would be much harder to identify 
	as they look like point sources with the correct PSF
	and cannot be matched to known asteroid databases. 
		
	For higher declination the rate of glints would be lower 
	than the rate presented here, 
	but is not well constrained by our survey. 
	It should be noted, that some high altitude satellites 
	follow high declination orbits (e.g., Tundra and Molniya orbits), 
	so the rate of flashes would be non-zero, 
	at least up to a declination of $\approx 65^\circ$. 
	In fact, \cite{satellite_glints_EvryScope_Corbett_2020} 
	report a non-zero rate of glints around the south celestial pole. 
	Since the astrophysical rate of short duration events is unconstrained, 
	even a very low rate of satellite glints at high declination could be 
	an important foreground. 
	
	For example, 
	while taking a spectrum of a high-redshift galaxy at a declination of $+62^\circ$, 
	\cite{GRB_flash_high_redshift_galaxy_Jiang_2020} 
	detected a point source that was seen in a single image.
	They suggested it was a GRB optical-UV emission. 
	The spectrum they present is not smeared, 
	implying the source was not streaked when crossing their slit,
	under $0.6''$ seeing conditions. 
	They therefore disqualify the possibility that the flash comes from a satellite.
	The flash could, however, appear point-like if it passed nearly perpendicular to the slit 
	\citep{GN_z_11_flash_satellite_Steinhardt_2021}
	or if its duration is shorter than 0.04\,s, 	
	which would be the case for an object rotating faster than once-per-minute
	\citep{GN_z_11_flash_satellite_Nir_2021}.  
	\cite{GRB_flash_high_redshift_galaxy_Jiang_2020} 
	estimate the flash has an equivalent magnitude of 19.2 in the V band. 
	Together with their exposure time of 179\,s we can estimate the native magnitude of a 0.04\,s glint
	would be $V\approx 10$, consistent with the magnitudes we observed in this work. 	
	The probability of having a satellite randomly cross their slit, 
	and produce a glint at the same time is low. 
	This, however, needs to be compared to the low probability of serendipitously detecting a GRB flash
	on the galaxy they were observing, 
	which is presumably orders of magnitude lower. 
	Finally, the GN-z11 flash was conclusively associated 
	with a specific spent booster rocket moving almost perpendicular 
	to the slit on a high-Earth orbit
	\citep{GN_z_11_flash_satellite_Michalowski_2021}.
	
	Thus, it is possible to confound astrophysical sources 
	with fast-rotating, high-altitude satellites, 
	in real world observations. 
	A dedicated survey for short-duration transients may therefore require
	a different strategy to mitigate this foreground. 
	
	One possible solution for ZTF could be to observe the same field
	with a smaller telescope (or set of telescopes) equipped with a fast CMOS camera.
	This auxiliary telescope system could cover a large part of the ZTF field of view, 
	with a cadence of 1\,s, reaching much shallower depth than the main ZTF camera, 
	but would be more sensitive to satellite glints because of its shorter exposure times.  
	If such a system is placed a few km from the main telescope, 
	satellite glints could be flagged based on their parallax between telescopes. 
	
	The Large Array Survey Telescope (LAST; \citealt{LAST_telescope_Ofek_Ben_Ami_2020}) 
	is another large area survey that will inevitably observe many satellite glints. 
	However, the survey strategy for LAST involves taking 
	a series of 20 exposures of 15\,s on the same field 
	and will be able to distinctly identify transients with durations longer than the exposure time. 
	In many cases, it could also detect multiple flashes in the same series of images, 
	and identify the source as an object moving in a straight line. 
	A similar strategy is proposed for LSST, 
	of taking two images back-to-back. 
	This strategy would help mitigate cosmic ray hits, 
	but could also be important for discriminating short transients 
	(on the order of the exposure time) from satellite glints. 
	Such strategies will still not be able to completely discriminate 
	glints from astrophysical transients that are shorter than the exposure time. 
	
	\section{Conclusions}\label{sec: conclusions}
	
	Since we have begun a blind search for fast transients in the W-FAST data, 
	we detected multiple instances of short duration flashes 
	lasting $\sim 0.2$\,s and reaching a typical brightness of 10-th magnitude. 
	While we cannot rule out that some of these flashes come from stellar flares
	of astrophysical origin, 
	we conclude that most or all of these events are satellite glints. 
	
	The size of the reflector required to produce such flares is cm-scale, 
	if the object is purely reflective. 
	If instead of a mirror we assume some transparent material like 
	glass covering optics or solar panels, 
	the size of the reflector would be on the 10\,cm scale. 
	
	In our sample we have many examples 
	where multiple flashes were seen from the same object. 
	In most cases the time delays between flashes are non-uniform, 
	and are much shorter than the rotation period we estimate
	based on the length of the flashes themselves. 
	This suggests such satellites have several reflective surfaces 
    of similar size and different orientations.
    
    The threshold we used to blindly detect such short flares
    in our data is very high compared to the noise, 
    suggesting that there are many other fainter, undetected glints.
    For individual flares the rate is $1.73^{+0.17}_{-0.16}$ and $1.26^{+0.17}_{-0.16}$ deg$^{-2}$ hour$^{-1}$  
    at declination $-25^\circ$ and $+5^\circ$, respectively. 
    For individual objects, counting all repeat flares as single objects, 
    we find a rate of $0.22^{+0.07}_{-0.05}$ and $0.07^{+0.05}_{-0.03}$ deg$^{-2}$ hour$^{-1}$ 
    for declinations $-25^\circ$ and $+5^\circ$, respectively. 
    This rate is an order of magnitude larger than the rate of glints reported by
    \cite{satellite_glints_EvryScope_Corbett_2020} 
    for the equatorial region. 
	One explanation for this difference
	is due to the fainter limiting magnitude of W-FAST. 
	Our exposures are shorter than the typical 
	flare duration, so the flares' flux is not diluted by the exposure time. 
	This allows us to observe fainter peak magnitudes, 
	and indeed we see many more flares in the 9--11\,mag range
	than flares brighter than 9\,mag
	(see Figure~\ref{fig: geosat magnitude histogram}). 
    
	Perhaps the most important conclusion is that 
	these flashes are quite common and could
	look similar to an astrophysical source
	in other surveys that search for short duration transients. 
	While satellites usually appear clearly streaked, 
	fast rotating objects may appear point-like. 
	
	While this survey hints that the rate of fast glints from
	geosynchronous satellites decreases at high declination 
	this cannot be taken as a rule. 
	Satellites have a wide variety of orbits, 
	and should always be considered as a foreground 
	to fast transient searches. 
	We discuss a possible observation of such a high declination glint, 
	reported by \cite{GRB_flash_high_redshift_galaxy_Jiang_2020} to be a GRB 
	in a high-redshift galaxy. 
	If the flash they report was, instead, a 0.04\,s duration satellite glint, 
	it would have been consistent with the satellites in our sample. 
	Eventually it was found that this flash was caused
	by a piece of space debris moving perpendicular 
	to the slit used in that observation
	\citep{GN_z_11_flash_satellite_Michalowski_2021}.
	This demonstrates the difficulty of discriminating short glints from astrophysical sources. 
	
	Finding real astrophysical short-duration transients
	requires developing ways to deal with this very high rate of foreground events. 
	One way to overcome this problem, 
	in an un-targeted search for astrophysical fast transients, 
	would be to scan only the area of the sky inside Earth's shadow. 
	For geosynchronous satellites this patch has a radius of $\approx 9^\circ$, 
	after accounting for the area lit by refracted light in Earth's atmosphere,
	and it becomes smaller for satellites on even higher orbits, 
	which are relatively few. 
	Another possibility is to view the same field using two telescopes
	placed $\gtrsim 1$\,km apart, so the parallax of flares can be measured.
	From our observations inside the Earth's shadow we can put an upper limit on 
	flares brighter than 11-th magnitude of $0.052$ deg$^{-2}$ hour$^{-1}$ (95\% confidence level). 
	
%	\pagebreak
	
	\section*{Acknowledgements}
	
	E.O.O.~is grateful for the support by grants from the 
	Israel Science Foundation, Minerva, Israeli Ministry of Science, Weizmann-UK, 
	the US-Israel Binational Science Foundation, 
	and the I-CORE Program of the Planning
	and Budgeting Committee.
	
	\section*{Data Availability}
	
	The data underlying this article will be shared on reasonable request to the corresponding author.
	
%	\clearpage
	
	\bibliographystyle{mnras}
	
	\bibliography{refs}
	
%	\vspace{10cm}
	
%	\clearpage
		
	\begin{deluxetable}{lcccccccc}
		
		\tabletypesize{\footnotesize}
		\tablewidth{14.5cm}
		
		\tablecaption{Observation log}
		
		\tablehead{
			\colhead{Run start time} & \colhead{RA}    & \colhead{Dec}   & \colhead{Duration} & \colhead{Altitude}  & \colhead{Airmass} & \colhead{Number}     & \colhead{Number}    & \colhead{In Earth's}  \\
			\colhead{(UTC)}          & \colhead{(Deg)} & \colhead{(Deg)} & \colhead{(hours)}  & \colhead{(degrees)} & \colhead{}        & \colhead{of objects} & \colhead{of flares}  & \colhead{shadow}
		}

		\startdata 
	
%	\begin{table*}		
%		\begin{tabular}{lccccccc}
	
	06-Aug-2020 18:23:49  &    272.17 & -20.41 &  0.6  &  39  &  1.59  &   0   &   0.0 &  No \\ 
	06-Aug-2020 19:02:54  &    305.50 & -17.47 &  2.0  &  40  &  1.55  &   2   &  81.0 &  No \\ 
	06-Aug-2020 21:16:34  &    334.30 & -10.49 &  2.0  &  48  &  1.35  &   2   &  45.0 &  No \\ 
	06-Aug-2020 23:32:25  &      10.45 & +5.96 &  2.0  &  63  &  1.12  &   0   &   0.0 &  No \\ 
	07-Aug-2020 01:47:40  &     58.00 & +22.00 &  1.0  &  59  &  1.16  &   0   &   0.0 &  No \\ 
	07-Aug-2020 20:24:09  &    331.00 & -10.49 &  1.0  &  42  &  1.49  &   0   &   0.0 &  No \\ 
	08-Aug-2020 21:01:43  &    331.00 & -10.50 &  0.3  &  44  &  1.43  &   0   &   0.0 &  No \\ 
	10-Aug-2020 19:30:37  &    272.17 & -20.41 &  0.1  &  37  &  1.67  &   0   &   0.0 &  No \\ 
	10-Aug-2020 19:55:51  &    272.16 & -20.42 &  1.1  &  32  &  1.90  &   2   &  14.0 &  No \\ 
	10-Aug-2020 21:06:47  &    272.16 & -20.42 &  0.0  &  27  &  2.22  &   0   &   0.0 &  No \\ 
	10-Aug-2020 17:54:39  &    305.50 & -17.50 &  0.9  &  31  &  1.93  &   1   &   7.0 &  No \\ 
	10-Aug-2020 21:13:56  &    337.74 & -10.49 &  2.0  &  48  &  1.35  &   0   &   0.0 &  No \\ 
	10-Aug-2020 23:40:42  &    331.00 & -10.50 &  0.0  &  45  &  1.41  &   0   &   0.0 &  No \\ 
	10-Aug-2020 23:56:06  &    331.09 & -10.63 &  2.0  &  35  &  1.73  &   0   &   0.0 &  No \\ 
	11-Aug-2020 02:14:20  &     98.47 & +21.49 &  0.6  &  31  &  1.94  &   0   &   0.0 &  No \\ 
	11-Aug-2020 19:49:54  &    305.62 & -17.49 &  0.4  &  41  &  1.52  &   0   &   0.0 &  No \\ 
	11-Aug-2020 17:40:57  &    275.00 & -25.00 &  0.6  &  34  &  1.80  &   0   &   0.0 &  No \\ 
	11-Aug-2020 19:22:53  &    305.49 & -17.50 &  0.4  &  40  &  1.57  &   0   &   0.0 &  No \\ 
	12-Aug-2020 18:52:41  &    272.17 & -20.41 &  0.1  &  38  &  1.61  &   0   &   0.0 &  No \\ 
	12-Aug-2020 19:05:14  &    272.30 & -20.40 &  0.9  &  36  &  1.68  &   3   &  25.0 &  No \\ 
	12-Aug-2020 17:50:36  &    305.49 & -17.52 &  0.9  &  32  &  1.91  &   0   &   0.0 &  No \\ 
	13-Aug-2020 20:41:23  &    331.02 & -10.49 &  0.6  &  45  &  1.41  &   0   &   0.0 &  No \\ 
	13-Aug-2020 21:17:35  &    331.04 & -10.49 &  1.3  &  49  &  1.33  &   2   &  44.0 &  No \\ 
	13-Aug-2020 22:48:47  &      10.50 & +6.01 &  1.3  &  59  &  1.17  &   0   &   0.0 &  No \\ 
	14-Aug-2020 00:20:44  &    330.99 & -10.54 &  1.4  &  32  &  1.88  &   0   &   0.0 &  No \\ 
	14-Aug-2020 02:19:57  &      10.51 & +6.00 &  0.5  &  55  &  1.21  &   0   &   0.0 &  No \\ 
	14-Aug-2020 17:23:29  &    272.17 & -20.40 &  1.2  &  39  &  1.59  &   5   &  48.0 &  No \\ 
	14-Aug-2020 18:52:22  &    272.15 & -20.42 &  2.0  &  34  &  1.79  &   9   &  34.0 &  No \\ 
	14-Aug-2020 21:01:21  &    331.03 & -10.49 &  1.5  &  49  &  1.33  &   3   &  75.0 &  No \\ 
	14-Aug-2020 22:46:49  &      10.50 & +6.01 &  1.3  &  59  &  1.17  &   0   &   0.0 &  No \\ 
	15-Aug-2020 00:15:42  &     29.00 & +17.02 &  2.0  &  73  &  1.05  &   1   &   3.0 &  No \\ 
	15-Aug-2020 02:33:11  &     29.00 & +17.00 &  0.2  &  74  &  1.04  &   0   &   0.0 &  No \\ 
	15-Aug-2020 17:15:41  &    272.17 & -20.39 &  1.3  &  39  &  1.59  &   6   &  53.0 &  No \\ 
	15-Aug-2020 18:44:27  &    272.15 & -20.43 &  2.0  &  34  &  1.77  &   1   &   2.0 &  No \\ 
	15-Aug-2020 21:05:47  &    331.01 & -10.48 &  1.4  &  49  &  1.33  &   0   &   0.0 &  No \\ 
	15-Aug-2020 22:35:17  &      10.50 & +6.01 &  1.4  &  58  &  1.17  &   0   &   0.0 &  No \\ 
	16-Aug-2020 00:20:16  &     29.41 & +17.02 &  2.0  &  74  &  1.04  &   0   &   0.0 &  No \\ 
	16-Aug-2020 02:35:23  &     29.00 & +17.00 &  0.3  &  73  &  1.05  &   0   &   0.0 &  No \\ 
	16-Aug-2020 18:43:36  &    272.15 & -20.43 &  2.0  &  34  &  1.79  &   3   &   8.0 &  No \\ 
	16-Aug-2020 17:40:56  &    305.52 & -17.50 &  0.9  &  32  &  1.86  &   7   &  24.0 &  No \\ 
	16-Aug-2020 21:03:52  &    331.01 & -10.48 &  1.3  &  49  &  1.33  &   1   &   2.0 &  No \\ 
	17-Aug-2020 18:54:56  &    272.18 & -20.42 &  2.0  &  32  &  1.86  &   3   &  10.0 &  No \\ 
	17-Aug-2020 17:25:52  &    305.50 & -17.49 &  1.4  &  33  &  1.84  &   2   &  12.0 &  No \\ 
	17-Aug-2020 21:09:10  &    305.50 & -17.50 &  1.1  &  37  &  1.64  &   0   &   0.0 &  No \\ 
	17-Aug-2020 22:38:35  &      10.40 & +5.91 &  1.3  &  59  &  1.16  &   0   &   0.0 &  No \\ 
	18-Aug-2020 00:06:24  &     29.00 & +17.02 &  2.0  &  73  &  1.04  &   0   &   0.0 &  No \\ 
	18-Aug-2020 02:23:14  &     29.00 & +17.00 &  0.5  &  72  &  1.05  &   0   &   0.0 &  No \\ 
	18-Aug-2020 19:10:12  &    272.15 & -20.41 &  1.3  &  33  &  1.86  &   0   &   0.0 &  No \\ 
	19-Aug-2020 18:47:56  &    272.15 & -20.43 &  1.9  &  32  &  1.86  &   0   &   0.0 &  No \\ 
	20-Aug-2020 00:07:49  &     29.00 & +17.02 &  1.9  &  74  &  1.04  &   0   &   0.0 &  No \\ 
	20-Aug-2020 02:36:21  &     58.00 & +22.00 &  0.3  &  75  &  1.03  &   0   &   0.0 &  No \\ 
	20-Aug-2020 18:39:54  &    272.15 & -20.43 &  2.0  &  33  &  1.85  &   0   &   0.0 &  No \\ 
	20-Aug-2020 17:37:14  &    305.50 & -17.49 &  0.9  &  34  &  1.79  &   5   &  56.0 &  No \\ 
	20-Aug-2020 20:52:37  &    342.07 & -10.51 &  2.0  &  48  &  1.35  &   0   &   0.0 &  No \\ 
	20-Aug-2020 23:06:26  &      10.50 & +6.01 &  0.6  &  62  &  1.14  &   1   &   2.0 &  No \\ 
	20-Aug-2020 23:50:13  &     29.01 & +17.02 &  2.0  &  73  &  1.05  &   0   &   0.0 &  No \\ 
	21-Aug-2020 02:08:17  &     29.00 & +17.00 &  0.7  &  72  &  1.05  &   0   &   0.0 &  No \\ 
	21-Aug-2020 17:20:07  &    305.50 & -17.49 &  2.0  &  36  &  1.69  &   0   &   0.0 &  No \\ 
	21-Aug-2020 19:33:50  &    331.00 & -10.49 &  1.4  &  44  &  1.44  &   0   &   0.0 &  Yes \\ 
	22-Aug-2020 18:02:14  &    272.16 & -20.42 &  1.5  &  37  &  1.67  &   2   &  79.0 &  No \\ 
	22-Aug-2020 21:16:17  &      10.50 & +6.02 &  2.0  &  53  &  1.26  &   2   &  17.0 &  No \\ 
	22-Aug-2020 23:28:09  &     29.01 & +17.00 &  0.1  &  60  &  1.16  &   0   &   0.0 &  No \\ 
	22-Aug-2020 23:41:40  &     29.00 & +17.02 &  2.0  &  73  &  1.05  &   0   &   0.0 &  No \\ 
	23-Aug-2020 18:09:26  &    272.16 & -20.42 &  0.9  &  37  &  1.65  &   0   &   0.0 &  No \\ 
	23-Aug-2020 19:10:58  &    331.00 & -10.48 &  2.0  &  44  &  1.43  &   0   &   0.0 &  Yes \\ 
	23-Aug-2020 21:24:35  &      10.51 & +6.03 &  2.0  &  55  &  1.22  &   0   &   0.0 &  No \\ 
	23-Aug-2020 23:37:13  &     28.81 & +16.83 &  2.0  &  72  &  1.05  &   0   &   0.0 &  No \\ 
	24-Aug-2020 01:54:04  &     29.00 & +16.99 &  1.1  &  71  &  1.06  &   0   &   0.0 &  No \\ 
	24-Aug-2020 20:43:26  &      10.50 & +6.04 &  2.0  &  48  &  1.34  &   0   &   0.0 &  No \\ 
	24-Aug-2020 23:11:56  &      10.50 & +6.01 &  0.2  &  63  &  1.12  &   1   &   6.0 &  No \\ 
	25-Aug-2020 00:01:34  &     29.01 & +17.01 &  0.1  &  68  &  1.08  &   0   &   0.0 &  No \\ 
	25-Aug-2020 00:27:22  &     29.05 & +17.01 &  1.3  &  76  &  1.03  &   0   &   0.0 &  No \\ 
	25-Aug-2020 01:58:55  &     29.06 & +16.99 &  1.0  &  70  &  1.07  &   0   &   0.0 &  No \\ 
	25-Aug-2020 17:14:22  &    294.35 & +38.96 &  2.0  &  77  &  1.02  &   0   &   0.0 &  No \\ 
	25-Aug-2020 19:33:50  &    331.00 & -10.48 &  2.0  &  47  &  1.37  &   0   &   0.0 &  Yes \\ 
	25-Aug-2020 21:44:56  &    331.00 & -10.50 &  1.6  &  46  &  1.39  &   0   &   0.0 &  Yes \\ 
	26-Aug-2020 17:13:05  &    294.36 & +38.95 &  1.9  &  78  &  1.02  &   1   &   2.0 &  No \\ 
	26-Aug-2020 20:27:42  &    331.01 & -10.48 &  1.2  &  49  &  1.33  &   0   &   0.0 &  Yes \\ 
	26-Aug-2020 21:57:25  &      10.51 & +6.01 &  1.4  &  59  &  1.17  &   2   &  80.0 &  No \\ 
	26-Aug-2020 23:42:22  &     30.73 & +16.82 &  2.0  &  74  &  1.04  &   0   &   0.0 &  No \\ 
	27-Aug-2020 01:58:44  &     29.00 & +16.99 &  1.0  &  68  &  1.08  &   0   &   0.0 &  No \\ 
	27-Aug-2020 19:51:46  &    294.34 & +38.93 &  1.9  &  65  &  1.11  &   0   &   0.0 &  No \\ 
	27-Aug-2020 18:26:18  &    331.00 & -10.49 &  1.2  &  37  &  1.65  &   0   &   0.0 &  No \\ 
	27-Aug-2020 21:55:35  &      10.50 & +6.01 &  1.0  &  57  &  1.19  &   1   &  17.0 &  No \\ 
	28-Aug-2020 17:15:05  &    294.36 & +38.95 &  1.2  &  76  &  1.03  &   0   &   0.0 &  No \\ 
	28-Aug-2020 19:49:58  &    294.36 & +38.94 &  0.6  &  72  &  1.05  &   0   &   0.0 &  No \\ 
	28-Aug-2020 18:48:28  &    333.50 & +51.51 &  0.9  &  58  &  1.17  &   0   &   0.0 &  No \\ 
	29-Aug-2020 17:06:25  &    294.35 & +38.95 &  1.3  &  75  &  1.03  &   0   &   0.0 &  No \\ 
	29-Aug-2020 21:01:47  &    294.36 & +38.94 &  1.0  &  55  &  1.22  &   0   &   0.0 &  No \\ 
	29-Aug-2020 22:04:05  &      10.31 & +5.92 &  2.0  &  63  &  1.12  &   2   &  45.0 &  No \\ 
	10-Sep-2020 17:05:36  &    272.16 & -20.42 &  1.2  &  36  &  1.69  &   1   &  10.0 &  No \\ 
	10-Sep-2020 19:20:46  &    331.00 & -10.50 &  0.2  &  47  &  1.37  &   0   &   0.0 &  No \\ 
	14-Sep-2020 18:05:49  &    272.16 & -20.42 &  0.9  &  29  &  2.07  &   1   &  12.0 &  No \\ 
	14-Sep-2020 16:48:16  &    305.50 & -17.49 &  1.2  &  40  &  1.57  &   0   &   0.0 &  No \\ 
	15-Sep-2020 16:49:07  &    272.17 & -20.41 &  1.2  &  36  &  1.70  &   0   &   0.0 &  No \\ 
	17-Sep-2020 19:06:27  &    305.50 & -17.50 &  0.6  &  39  &  1.58  &   0   &   0.0 &  No \\ 
	18-Sep-2020 16:25:21  &    272.16 & -20.43 &  1.6  &  36  &  1.71  &   4   &  18.0 &  No \\

		\vspace{0.2cm}
		\enddata
		
		\vspace{-0.5cm}
		\tablecomments{
%		\end{tabular}
%		\caption{
			The coordinates are for the centre of the field for that run (in J2000). 
			Also specified is the run duration. 
			The airmass and altitude are taken for the middle of the run. 
			Number of objects is the number of distinct objects, possibly with multiple flares, seen in that run. 
			Number of flares refers to the total number of separate flares from all objects in that run. 
			We define a field to be `in Earth's shadow' if the line of sight of the centre of the field 
			is inside Earth's shadow when crossing the height of geosynchronous satellites ($\sim 37000$\,km). 
			Such fields are not expected to have any contributions from any satellites up to (and including) 
			geosynchronous orbit. 
		}
		
		\label{tab: observation log}
		
	\end{deluxetable}
%	\end{table*}
\bsp	% typesetting comment
\label{lastpage}
\end{document}